\documentclass[12pt]{article}

\usepackage{graphicx}
\usepackage{epsfig}
\usepackage[dvips]{color}

\newcommand{\be}{\begin{eqnarray}}
\newcommand{\ee}{\end{eqnarray}}

\begin{document}


\thispagestyle{empty}
\setcounter{page}{1}

\hbox{} \nopagebreak \vspace{-3cm}
\addtolength{\baselineskip}{.8mm} \baselineskip=24pt


\begin{flushright}
{\sc BI-TP 2004/02}\\
\end{flushright}
\vspace{-1.0cm}
\begin{flushright}
{\sc CU-TP-1107}\\
\end{flushright}
\vspace{-1.0cm}
\begin{flushright}
{\sc LTP-Orsay 04-23 }\\
\end{flushright}

\vspace{40pt}

\begin{center}
{\large {\sc {\bf  Saturation and shadowing in high-energy\\
proton-nucleus dilepton production}}}\\
\baselineskip=12pt \vspace{34pt}
R.~Baier $^a$, A.H.~Mueller $^b$, D.~Schiff $^c$
\vspace{24pt}

$^a$ Fakult\"at f\"ur Physik, Universit\"at Bielefeld, 
D-33501 Bielefeld, Germany
\vspace{7pt}

$^b$ Department of Physics, Columbia University, New York, NY 10027, USA
\vspace{7pt}

$^c$ LPT, Universit\'e Paris-Sud, B\^atiment 210, F-91405 Orsay, France

\vspace{60pt}

\end{center}

\vspace{40pt}

\begin{abstract}

We discuss the inclusive dilepton cross section for 
proton (quark)-nucleus collisions at high energies
in the very forward rapidity region. Starting from the calculation 
in the quasi-classical approximation, 
we include low-$x$ evolution effects in the nucleus and predict
leading twist shadowing together with anomalous scaling behaviour.

\end{abstract}

\vfill

\newpage


\section{Introduction}

There is increasing evidence that hard probes \cite{Accardi} 
are an excellent tool for analyzing
the matter produced in high-energy heavy ion collisions
at RHIC \cite{Ludlam}, especially when 
calibrated against similar probes in proton-proton and proton (deuteron)-ion
reactions \cite{Wiedemann}.
 At central rapidities the fact that high-$p_\bot$ hadron production 
in $AA$ reactions is suppressed as compared to the production 
in $pp$ collisions
times the expected number of hard collisions,
 $N_{coll}$, gave strong support to
the idea that dense, hot matter is produced in $AA$ collisions causing jets to 
loose a significant amount of their energy, while passing through the dense 
matter and before producing the observed high-$p_\bot$ hadron. This picture was
further confirmed when, at central rapidities,
 high-$p_\bot$ hadron production in 
$dA$ collisions did not show any suppression as
 compared to the expectation from 
$pp$. The lack of a suppression in $dA$ as compared to
 $pp$ collisions of course
also means that there is little or no nuclear shadowing, at the hard scale 
determined by high-$p_\bot$ hadron production, 
in the central rapidity region \cite{Back}.

Recent $dA$ data on high-$p_\bot$ hadron
 production at large rapidity (toward the
deuteron side) from the BRAHMS Collaboration \cite{Debbe} 
show a significant suppression of
hadron production in $dA$ collisions compared to the expectation from $pp$
collisions. 
This result has aroused a lot of interest,
 because it suggests that there may be
a significant amount of (leading twist) gluon shadowing in nuclear 
wavefunctions in the region probed by forward hard scattering at RHIC. The 
strong interest is connected to the fact that strong (leading twist) gluon 
shadowing appears difficult to understand outside of pictures which have  
gluon saturation \cite{GLR,Mueller:wy}
(color glass condensate \cite{MV,Iancu:2003xm,Dumitru}),
 which to a significant extent 
is driven by BFKL evolution \cite{BFKL}. 

In many ways hard photon or $\mu$-pairs coming from virtual
 photons\cite{Arleo,Fai} are a better
probe than high-$p_\bot$ hadrons
\cite{KoM,Kovch,YKdif,KNST,KLM,bkw,KW,cronin,JNV,
Blaizot:2004,JJM:2004,Iancu:2004}.
 With hard photons \cite{Brodsky,Kopeliovich,Raufeisen,GJM,Gelis:2002ki,JMJ}
 one is less sensitive 
to fragmentation effects and final state effects are absent. This 
means that at transverse momenta around $2 - 3$ GeV, 
one can expect leading twist 
factorization to be accurate, and hence $x$-values of 
the gluon distribution of 
the nucleus down to values somewhat smaller than $10^{-3}$ should be
accessible. The main purpose of this paper is to explore,
 and estimate, the size
of the suppression one might expect to see in such reactions. Our discussion
is based on a picture, where the McLerran-Venugopalan model \cite{MV}
 is taken to 
represent the gluon distribution in a hard RHIC reaction at central values 
of rapidity, $y=0$, and BFKL evolution \cite{BFKL,MT} 
is used to evolve the distribution to higher values of $y$. 

There has already been quite a lot of work studying hard photon and
$\mu$-pair production in $p(d)A$ collisions \cite{Brodsky}.
 In a pioneering series of 
papers Kopeliovich and collaborators \cite{Kopeliovich,Raufeisen}
 have studied Drell-Yan production in the
RHIC and LHC kinematic regions using a dipole picture of the $\mu$-pair
production. Gelis and Jalilian-Marian \cite{GJM,Gelis:2002ki}
arrived at equivalent results in a color 
glass condensate picture, where the dipole of Kopeliovich et al. is 
replaced by a product of two Wilson lines evaluated in the field of the 
color glass condensate. Jalilian-Marian \cite{JMJ}
then calculated the suppression factor, however, for the
 $k_\bot$-integrated yields
in $\mu$-pair production in $p(d)A$ versus $pp$ reactions taking the dipole 
cross section as determined by Iancu, Itakura and Munier \cite{IIMu}
in fits to HERA
data, based on the geometric scaling following from BFKL dynamics not too far
from the saturation boundary of the color glass condensate. 
Quite a strong suppression is found in the analysis of \cite{JMJ}, 
because the gluon distribution used there has leading twist 
shadowing in contrast to the models in
 \cite{Kopeliovich,Raufeisen,GJM,Gelis:2002ki}

In this paper we evaluate direct photon and $\mu$-production in terms of 
standard factorization formulae.
 We remind the reader, how $k_\bot$-factorization
formulae arise, and why $k_\bot$-factorization is more efficient than ordinary 
operator product factorization, when one is dealing with small-$x$ processes. 
Our general discussion is not tied to saturation or color glass condensate 
assumptions, but rather is a general leading twist discussion. However, 
because it is leading twist only, in contrast to previous discussions, it 
should only be used for moderate transverse momentum, say $k^2_\bot \ge 4
{~\rm GeV}^2$. When we take the unintegrated gluon distribution, which appears
in our formulation to
 be given in terms of the anomalous scaling, which occurs in 
the BFKL based saturation picture \cite{MT},
 our overall picture is very
close to \cite{JMJ}. 

The outline of our paper is as 
follows:

In Sec.~2, we derive a $k_\bot$-factorized formula for high-$k_\bot$ 
transversely polarized (virtual) photons produced in a quark-nucleus 
(hadron) collision. The corresponding formula for lepton-pair production, 
with lepton pair mass $M$, is  given by
\begin{equation}
\frac{d\sigma^{qA\rightarrow l^+ l^- X}}{d^2 b d^2 k_\bot d \ln z d M^2} = 
\frac{\alpha_{em}} {3 \pi M^2} \frac{d\sigma^{qA\rightarrow \gamma^* X}}{d^2
b d^2 k_\bot d \ln z} \, , 
\label{(1)}
\end{equation}
where $k_\bot$ is the transverse momentum of the $\gamma^*$, and $z$ is
the longitudinal momentum fraction of the $\gamma^*$
 with respect to the incident
quark momentum, $z=k_+ / p_+$, where we have the limit $p_+ \rightarrow \infty$
in mind.
 $\vec b$ denotes the impact parameter of the
$qA$ collision. 
Throughout the paper we shall refer to direct 
photon production, but lepton-pair production formualae follow easily from 
(\ref{(1)}). 
For simplicity we consider incident quarks rather than protons. 

In Sec.~3, we review the form that 
the unintegrated gluon  distribution takes in 
saturation (color glass condensate) models. We do this first in the 
McLerran-Venugopalan model \cite{MV}, which has gluon saturation but no gluon 
shadowing, and then for the case, where a significant amount of BFKL evolution 
is added to the McLerran-Venugopalan model,
 which is taken as the initial 
condition for that evolution.
 With BFKL evolution \cite{BFKL,MT} gluon shadowing appears and the 
fixed impact parameter unintegrated gluon distribution scales with $A$
(roughly) like $A^{\frac{(1-\lambda_0)}{3}}$, with $\lambda_0 \simeq 0.372$.

In Sec.~4, we present numerical results, which suggest a significant 
suppression of hard photons in the forward rapidity region in $p(d)A$
collisions as compared to $pp$ collisions. Our results,
we hope, are encouraging for 
experimenters trying to measure the suppression at RHIC.  

In Appendix A we relate the $k_\bot$- to the impact
parameter representation.

In Appendix B and C we relate our $k_\bot$-factorized formulation to the more 
standard QCD factorization.
We show explicitly that the anomalous scaling formulae, which appear in the  
$k_\bot$-factorization formalism, lead to an (integrated) gluon distribution
which obeys the renormalization group equation with an anomalous 
dimension given by BFKL evolution.

\section{Dilepton production cross section}

\subsection{Factorized formula for the inclusive $\gamma^* $ cross section}

In the following we only consider the case for virtual photons with transverse 
polarizations, $\lambda = 1 ,2$. 
The $qA\rightarrow \gamma^* X$ cross section is obtained from the
$k_{\bot}$-factorized formula 
 containing the photon phase space, the photon emission amplitude 
$A_\lambda$ and the unintegrated gluon distribution $\phi_G$, explicitly
\begin{equation}
d\sigma^{qA \rightarrow \gamma^* X} = \frac{d^3 k}{2 k^0 (2\pi)^3} 
\, \int \, \frac{g^2}{
2N_c} \phi_G (\vec q_\bot , Y) \frac{d^2 q_\bot}{q^2_\bot} \sum_\lambda 
| A_\lambda |^2 , 
\label{(2)}
\end{equation}
where
$\frac{g^2}{2 N_c} \frac{\phi_G (\vec q_\bot , Y)}{ q^2_\bot}$
may be viewed as the differential high energy
 $qA \rightarrow qA$ cross section.
In the definition of $\phi_G$ in (\ref{(2)})
 the integration with respect to the impact parameter 
is implied, 
\begin{equation}
\phi_G (\vec q_\bot , Y) \equiv \int d^2 b \phi_G (\vec b , \vec q_\bot , 
Y = \ln 1/x) . 
\label{(6)}
\end{equation}
 We shall discuss  in the 
following subsection, how $k_\bot$-factorization applies to
$qA \rightarrow \gamma^* X$ at large  $k_\bot$, and how it leads to
(\ref{(2)}) (see also \cite{BDMPS}).

The photon emission amplitude $A_\lambda$ is expressed in terms of the 
polarization vector $\vec\epsilon^{~\lambda}_\bot$ and the 
transverse momenta $\vec k_\bot$, and $\vec k^{~\prime}_\bot
=\vec k_\bot - z \vec q_\bot$, respectively as 
\begin{equation}
A_\lambda = - 2 ie \left( \frac{\vec\epsilon^{~\lambda}_\bot \cdot \{ 
\vec k_\bot \} } {k^2_\bot + \eta^2} - \frac{ \vec\epsilon^{~\lambda}_\bot 
\cdot \{ \vec k^{~\prime}_\bot \} }{k^{\prime 2}_\bot + \eta^2} \right) , 
\label{(7)}
\end{equation}
where we use the short hand notation \cite{Baier}
\begin{equation}
\vec\epsilon^{~\lambda}_\bot \cdot 
\{ \vec k_\bot \} = (1 - z/2 ) \vec\epsilon^
{~\lambda}_\bot \cdot \vec k_\bot - i z/2 ~\vec\epsilon^{~\lambda}_\bot 
\wedge \vec k_\bot , 
\label{(8)}
\end{equation}
and 
\begin{equation}
\eta^2 = (1 - z ) M^2 . 
\label{(9)}
\end{equation}
Performing the polarization sum we note that 
\begin{equation}
\sum_\lambda | A_\lambda |^2 = \frac{1 + (1 -z)^2}{2} \sum_\lambda 
(4e^2 ) \left( \frac{\vec\epsilon^{~\lambda}_\bot \cdot \vec k_\bot}{k^2_\bot 
+ \eta^2} - \frac{\vec\epsilon^{~\lambda}_\bot \cdot \vec k^{~\prime}_\bot}
{k_\bot^{\prime 2} + \eta^2} \right)^2 . 
\label{(10)}
\end{equation}
Finally the transverse $\gamma^*$ production cross section becomes 
\begin{eqnarray}
& &\frac{d\sigma^{qA \rightarrow \gamma^* X}}{d^2 b d^2 k_\bot d \ln z}  =  
\frac{\alpha_{em} \alpha_s}{\pi N_c} [ 1 + (1-z)^2] \int \frac{d^2 q_\bot}{
q^2_\bot} \phi_G (\vec b , \vec q_\bot , Y ) \label{(11)} \\
& & \times \left\{ \frac{z^2 \vec q^{~2}_\bot}{(k^2_\bot + \eta^2) [(
\vec k_\bot - z \vec q_\bot )^2 + \eta^2 ]} - \eta^2 
\left[ \frac{1}{k^2_\bot + \eta^2} - \frac{1}{(\vec k_\bot - z \vec q_\bot)^2 
+ \eta^2} \right]^2 \right\} .   \nonumber 
\end{eqnarray}
We remark that in the paper \cite{GJM}
the longitudinal contribution is included\footnote{The function $C (l_\bot)$ 
introduced in \cite{GJM}
corresponds to $\frac{(2\pi)^3 \alpha_s}{N_c} \frac{\int d^2 b \phi_G 
(\vec b , \vec l_\bot)/l^2_\bot}{\pi R^2}$.}
\begin{eqnarray}
\frac{d\sigma^{qA\rightarrow \gamma^* X}}{d^2 b d^2 k_\bot d \ln z} & = & 
\frac{\alpha_{em} \alpha_s}{\pi N_c} 2 (1-z) \int \frac{d^2 q_\bot}{q^2_\bot}
\phi_G (\vec b , \vec q_\bot , Y) \nonumber \\
& & \times \eta^2 \left[ \frac{1}{k^2_\bot + 
\eta^2} - \frac{1}{(\vec k_\bot - z \vec q_\bot )^2 + \eta^2} \right]^2 . 
\label{(12)}
\end{eqnarray}
In Appendix A we express the $k_\bot$-factorized cross section 
(\ref{(11)}) in terms of the dipole formulation in the impact 
parameter representation \cite{Brodsky, Kopeliovich, Raufeisen}  
and we shortly mention the relation to DIS.
 Appendix B discusses the relation of 
(\ref{(11)}) to the large $k_\bot$ LO pQCD cross section for the case 
of real (isolated) photons.

\subsection{Hard reactions and $k_\bot -$ factorisation}

Before we continue and discuss the evolution of $\phi_G$ with respect to
increasing rapidity $Y$ we briefly recount the origin and role
of $k_{\bot}$-factorization in small-$x$ hard reactions,
especially the validity of Eq.(\ref{(2)}).

In the usual QCD factorization \cite{Ellis} involving
 local gauge invariant operators
in the operator product expansion at small-$x$ it may be necessary to 
resum $\alpha_s \ln 1/x$ terms in both the coefficient functions and in the 
matrix elements, evaluated at a hard scale $Q^2$, which occur. There is an 
alternative procedure in which the hard part of the reaction can be taken 
at lowest order in perturbation theory and the resummation done on what 
remains. In this $k_\bot$-factorization a convolution in transverse momentum 
then remains to be done between the ``factorized'' parts while there is no 
convolution in longitudinal momentum because the formalism only exists in a 
leading order formulation. Indeed one of the shortcomings of the 
$k_\bot$-factorization formalism is that it is not known whether this leading 
order procedure is part of a more systematic procedure or not. On the other 
hand $k_\bot$-factorization is very useful when extremely small values of 
$x$ are being considered where resummations in $\alpha_s \ln 1/x$ are 
paramount, which resummations are somewhat awkward in the standard hard 
QCD factorization \cite{Ellis}. 

Let us now examine how $k_\bot$-factorization comes about in the process 
of interest here, direct photon production in, say, quark-nucleon (or nucleus)
collisions. The process is illustrated for a ``typical'' graph in 
Fig.~1.

\begin{figure}[htb]
\begin{center}
\epsfig{bbllx=0,bblly=0,bburx=280,bbury=160,
file=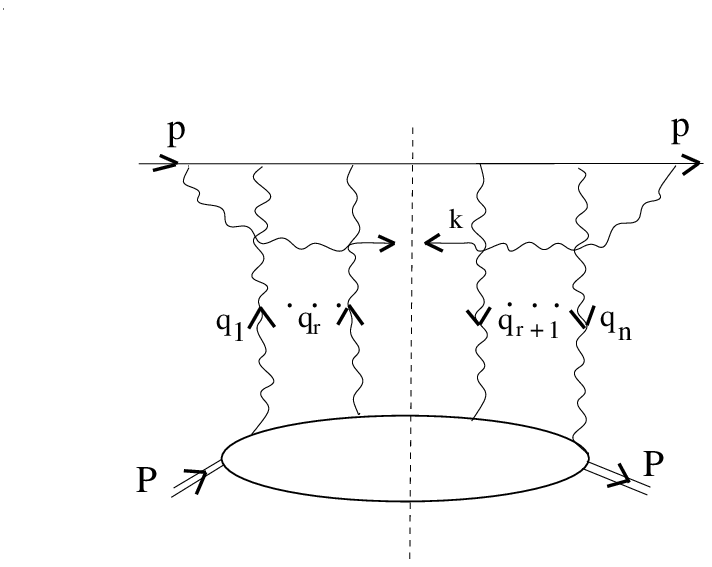,width=10cm}
\caption{Typical graph for photon production
in quark-nucleon scattering.}
\label{fig1}
\end{center}
\end{figure}


\noindent 
$p$ is the incoming quark, $P$ the target and $k$ the hard photon setting 
the hard scale for the process. The lines $q_1, ... q_r$ are gluons 
exchanged in the amplitude while $q_{r+1}, ... q_n$ are the ones 
in the complex conjugate amplitude. 
We suppose that $p_+$ and $k_+$ are large, with $z = k_+ / p_+$ fixed, and 
we further suppose that all the gluons and quarks in the lower ``blob''
of the graph have $+$ components of the momentum much less than $k_+$. 
This latter assumption is important in $k_\bot$-factorization; a strong 
ordering in longitudinal momentum is necessary. In addition we suppose that 
$k_\bot$ is large while, for simplicity of discussion we take 
$p_\bot = 0$. The lines, $q_i$, which connect the hard part of the graph
with the target, $P$, in general have $q_{i +} \ll k_+ , p_+$. Now we limit
our discussion to leading twist, in $k_\bot^2$. In a covariant gauge there 
may be many $q_i$-gluons present, however in light cone gauge, with 
$A_- = 0$, the leading twist contribution can involve only two 
exchanged gluons \cite{KoM,Kovch}
 Thus, taking $A_- = 0$ we consider the graph shown in Fig.~2. 

\begin{figure}[htb]
\begin{center}
\epsfig{bbllx=0,bblly=0,bburx=340,bbury=170,
file=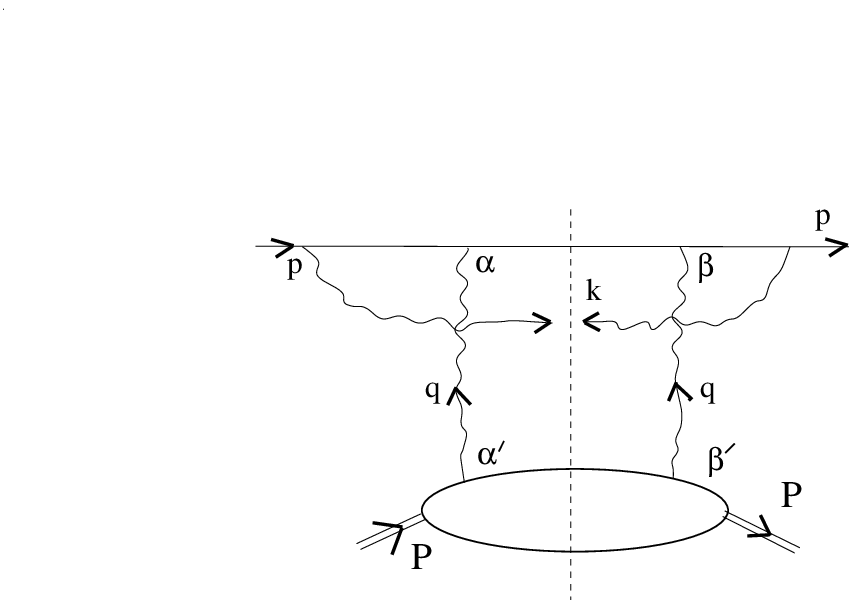,width=11cm}
\caption{Two gluon exchange graph. }
\label{fig2}
\end{center}
\end{figure}


\noindent
There are three other graphs in the two gluon exchange or leading twist 
limit, however, they may be ignored as our object here is to explain 
$k_\bot$-factorization not to give a detailed calculation of terms in the 
factorized formula.

In $A_- = 0$ light cone gauge the propagator is 
\begin{equation}
D_{\alpha \alpha^\prime} (q) = \frac{-i}{q^2} \left[ g_{\alpha \alpha^\prime}
- \frac{\bar\eta_\alpha q_{\alpha^\prime} + \bar\eta_{\alpha^\prime} q_\alpha}{
\bar \eta \cdot q} \right] , 
\label{(4.01)}
\end{equation}
with $\bar\eta \cdot v = v_-$ for any vector $v_\mu$.
 When applied to the hard part of the graph shown in Fig.~2, 
or to any of the other possible hard parts that may occur, 
it becomes \cite{Mueller:2001fv}
\begin{equation}
D_{\alpha \alpha^\prime} (q) \rightarrow \frac{i}{q^{~2}_\bot} \bar\eta_{
\alpha^\prime} \eta_\alpha \, , 
\label{(4.3)}
\end{equation}
and similarly for $D_{\beta\beta^\prime} (q)$, with
$\eta \cdot v = v_+$.
The $q_\alpha$ term in (\ref{(4.01)}) 
gives zero by current conservation, while $\bar\eta_\alpha$ 
projects a relatively
small component of the momenta.
 This leaves only the $g_{+-}$ term
in (\ref{(4.01)}), which corresponds to (\ref{(4.3)}).

While (\ref{(4.3)}) looks like a covariant gauge result this is not quite the 
case. In covariant gauge the leading twist contribution is not limited 
to the two gluon exchange term shown in Fig.~2; 
the many gluon exchange terms of Fig.~1 are also important. In addition, in the
present case, the lower blob in Fig.~2 must be evaluated in light cone gauge. 
A covariant gauge evaluation will give an incorrect result. 

Taking the graph shown in Fig.~2 along with the three graphs where either, or
both, of the exchanged $q$-lines hook into the hard part of 
the graph before the 
photon, $k$, is emitted leads to the cross section formula given in Sec.~2. 
The unitegrated gluon distribution $\phi_G (\vec q_\bot , Y)$ is given by 
$\frac{\bar\eta_{\alpha^\prime}  \bar\eta_{\beta^\prime}}{q^{~2}_\bot} 
\frac{d q_+}{(2\pi)^3}$ acting on the graphs in the lower blob of Fig.~2, as 
we now illustrate in Fig.~3.

\begin{figure}[htb]
\begin{center}
\epsfig{bbllx=20,bblly=90,bburx= 320,bbury=175,
file=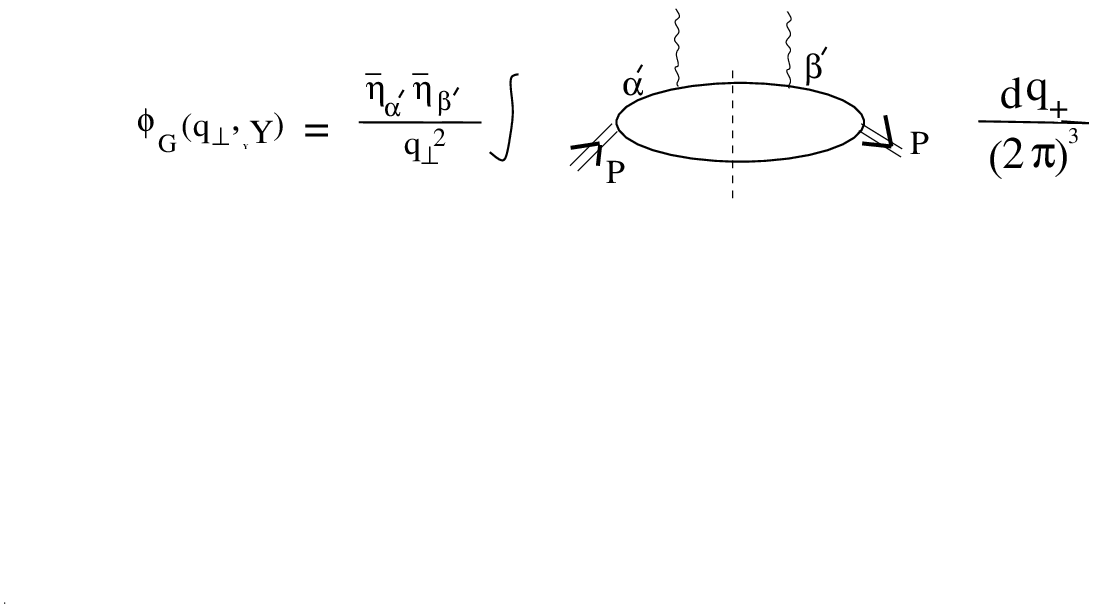,width=13cm}
\caption{Unintegrated gluon distribution. }
\label{fig3}
\end{center}
\end{figure}


\noindent
Our normalization is such that 
\begin{equation}
\phi^{(0)}_G (\vec q_\bot , Y) = \frac{\alpha_s C_F}{\pi}
 \frac{1}{q^2_{\bot}}  
\label{(4.4)}
\end{equation}
for a quark at lowest order in $\alpha_s$ \cite{Mueller:1999wm},
 and an additional 
factor $N_c$ for three quarks in a nucleon gives finally, e.g. (\ref{(3.5)}). 

For large $Q^2$ the following relation, using the $qA \rightarrow qA$
cross section,
between $\phi_G$ and the gluon structure function may be  derived \cite{BDMPS},
\begin{eqnarray}
\int^{Q^2}_0 d^2 q_\bot q^2_{\bot} \frac{d\sigma^{qA}}{d^2  q_\bot} 
& = & \frac{g^2}{2 N_c} \int \phi_G (\vec q_\bot , Y) d^2 q_\bot \nonumber
\\ \label{(4)} \\
& = & \frac{4\pi^2 \alpha_s C_F}{N^2_c - 1} x G_A (x , Q^2) \, ,
\nonumber 
\end{eqnarray}
 which, however, cannot be used in the scaling region.

In contrast to normal hard QCD factorization $k_\bot$-factorization requires a 
convolution in transverse momentum be taken between the hard part and the 
unintegrated gluon distribution to
arrive at a cross section, as given for example in (\ref{(2)}). 

\section{Saturation and anomalous scaling}

In the leading twist region
the function $ \phi_G (\vec b , \vec q_\bot , Y )$  in (\ref{(6)}),
the unintegrated gluon structure function\footnote{This function
is sometimes denoted as modified gluon distribution $h(\vec q_\bot, Y)$
\cite{KW,Armesto:vs}.},
 is expressed in terms of the forward scattering amplitude
$N(\vec b , \vec x_\bot , Y ) = N_{q \bar q}(\vec b , \vec x_ \bot , Y )$
of a QCD $q \bar q$ dipole of transverse size $\vec x_ \bot$ with
rapidity $Y=\ln{1/x}$, scattering off a nucleus $A$ at impact parameter
$\vec b$, by

\begin{eqnarray}
 \phi_G (\vec b , \vec q_\bot , Y ) &=& \frac{N_c}{(2\pi)^3 \alpha_s} 
\int d^2 x_\bot  e^{i \vec q_\bot \cdot \vec x_\bot}
 \vec\nabla^2_{x_\bot} N(\vec b , \vec x_\bot , Y) 
 \nonumber \\ \label{(III.1)} \\ 
 &=&\frac{N_c}{(2\pi)^3 \alpha_s}\,   q^2_{\bot} \vec\nabla^2_{q_\bot}
 \int \frac{d^2 x_\bot}{x^2_{\bot}}  e^{i \vec q_\bot \cdot \vec x_\bot}
N(\vec b , \vec x_\bot , Y) \, .
\nonumber 
\end{eqnarray}
The function  $\phi_G (\vec b , \vec q_\bot , Y )$ may also be expressed
in terms of the forward amplitude of a gluon-gluon dipole
$ N_G(\vec b , \vec x_\bot , Y)$, obtained by replacing the
 number of colors $N_c$
by $C_F$ \cite{cronin,KTin}.
The relation (\ref{(III.1)}) is inverted by
\begin{equation}
 N(\vec b , \vec x_\bot , Y) = \frac{(2\pi)^3 \alpha_s}{2 N_c} 
\int \frac{d^2 q_\bot}{(2\pi)^2  q^2_{\bot}} \, 
\phi_G (\vec b , \vec q_\bot , Y )
[ 2 -  e^{-i \vec q_\bot \cdot \vec x_\bot} 
- e^{i \vec q_\bot \cdot \vec x_\bot} ] \, ,
\label{(III.2)}
\end{equation}
illustrating the relation  of $\phi_G$ to the
 $qA \rightarrow qA$ cross section 
more directly. The phase factors in the bracket are due to the four graphs
describing the different ways of gluon exchanges of
the $q \bar q$ scatterings of the target nucleus \cite{Patel}.

Let us emphasize that the unintegrated gluon distribution $\phi_G$ as
defined in the previous section (see Fig.~3) is an object to be used only in
leading twist $k_\bot$-factorized formulae.
Eqs. (\ref{(III.1)}) and (\ref{(III.2)}) are leading twist equations,
valid in the scaling region - which we shall discuss later -
and beyond.

 For illustration and later reference, we shortly summarize in Appendix B
the pQCD leading order (LO) behaviour of  $N$ and $\phi_G$.

\subsection{McLerran-Venugopalan model}

Let us review the quasi-classical  model by McLerran-Venugopalan \cite{MV}
(at fixed $\vec b$ and at $Y=0$)
as a reasonable starting point of the $Y$ evolution of $\phi_G$,
\begin{equation}
N^{MV} (\vec b, \vec x_\bot , Y=0) = 1 - \exp [ - x^2_\bot \bar Q^{~2}_s (
\vec b ) /4],
\label{(3.10)}
\end{equation}
with the 
 saturation scale \cite{Mueller:2001fv,Mueller:1999wm} given by
\begin{equation}
\bar Q^{~2}_s (\vec b) = 
\frac{2 \pi^2 \alpha_s}{N_c}
\rho T (b) (x G (x , 1/ x^2_\bot ) ) \, ,
\label{(3.7)}
\end{equation}
with $\rho$ the nuclear density and $T(b)$ the profile function $T(b) = 
2 \sqrt{R^2_A - b^2}$. $xG$ is the gluon distribution in the nucleon.
The unintegrated gluon 
distribution $\phi^{MV}_G (\vec b , \vec q_\bot , Y=0)$, calculated from 
(\ref{(III.1)}), approaches $\phi^{\rm{LO}}_G$ (\ref{(3.5)}) at large $\vec q_\bot$, 
i.e. for $q_\bot \gg \bar Q_s$, from above. The low momentum part is 
suppressed relative to the perturbative gluon; keeping $\bar Q_s$ constant, 
independent on $x_\bot$, one derives in this model for $q_\bot \ll \bar Q_s$, 
\begin{equation}
\phi^{MV}_G (\vec b, \vec q_\bot , Y = 0) \simeq \frac{N_c}{2 \pi^2 \alpha_s}
\frac{q^2_\bot}{{\bar Q}^2_s} .
\label{(3.11)}
\end{equation}

As discussed in some detail in \cite{cronin},
 it may
nevertheless be worthwhile  to note that
this behaviour differs from the one known from the unintegrated gluon distribution
derived from the non-Abelian Weizs{\"a}cker-Williams field of a nucleus, denoting it 
by $\phi^{WW}$,
which behaves in the quasi-classical approximation as 

\begin{equation}
 \phi^{WW} (\vec b, \vec q_\bot, Y=0) \simeq \frac{N_c}{2 \pi^2 \alpha_s}
 \ln{(\frac{\bar Q_s}{q_\bot})^2} \, .
\label{(18p)}
\end{equation}
Both functions, $\phi_G$ and $\phi^{WW}$, do, however, agree for $q_\bot >> \bar Q_s$.
When $q_\bot << \bar Q_s$, all twists become important in this kinematic regime,
 and indeed, comparing (\ref{(3.11)}) and (\ref{(18p)})
    there is no unique prescription to define the 
unintegrated gluon distribution. In order to
calculate the photon spectrum (\ref{(2)})
  at $Y=0$ we continue to use $\phi_G$, as defined by
(\ref{(III.1)}),
 together with (\ref{(3.10)}),
knowing that we will be interested in large enough transverse momentum only,
so that in fact $\phi_G \sim \phi^{WW}$.

The region around $q_\bot \simeq \bar Q_s$ is enhanced, since the effect of 
multiple scatterings, resummed in (\ref{(3.10)}), rearrange the gluons in the 
nucleus \cite{Dumitru, bkw,cronin}. 
There is no shadowing in the quasi-classical approximation. 

Starting with (\ref{(3.10)}) already at RHIC energies for dileptons produced
in the central region $(y_\gamma \simeq 0)$ implies that the $x$-values 
in the gluon function in the nucleus are already small enough in order to 
justify the applicability of the McLerran-Venugopalan model, although 
there is no explicit $x$-dependence in this model. 
A rough estimate, based on hard two partons $\rightarrow \gamma^*$ 
$k_\bot$-factorized kinematics, gives $x \simeq \frac{M_\bot}{\sqrt s}$ 
at central rapidity; for collision energy $\sqrt s = 200$ GeV
 and a transverse mass 
$M_\bot = 4$ GeV a reasonable small value of $x \simeq 0.02$
follows. This 
implies that the gluon number density at RHIC energies
\cite{Ludlam} is already large, i.e. 
saturated \cite{GLR,Mueller:wy}, 

\begin{equation}
\int \phi^{MV}_G (\vec b , \vec q_\bot , Y=0 )d^2
 q_\bot \simeq \frac{N_c \bar Q^{~2}_s (\vec b)}{2\pi \alpha_s}
\label{(3.12)}
\end{equation}
for fixed $\bar Q_s (\vec b)$.

\subsection{BFKL evolution in the presence of saturation}

Increasing the photon rapidity into the forward region, $y_\gamma > 0$, the 
values of $x$ become rapidly small, namely $x \simeq (M_\bot / \sqrt s ) 
e^{-y_\gamma}$, such that $Y=\ln 1/x \simeq y_\gamma$ increases with 
$y_\gamma$.
In the following we fix $Y=0$ at $y_\gamma = 0$ and treat $Y$
as equivalent to $y_\gamma$ for positive large rapidities. 
We work with the fixed coupling leading order approximation of the 
$Y$ evolution, which effectively depends on the product of $\alpha_s Y$.  

In order to calculate the $Y$ dependence of $\phi_G$ we start from the 
BFKL evolution \cite{BFKL}
and write the amplitude $N (\vec b, \vec x_\bot , Y)$ in terms of the 
Mellin transform
\begin{equation}
N (\vec b , \vec x_\bot , Y) = - \int \frac{d\lambda}{2\pi i} \Gamma 
(\lambda - 1) \exp \left[ 2 \bar\alpha \chi (\lambda ) Y + 
(1 - \lambda ) \ln (x^2_\bot \bar Q^{~2}_s (\vec b)/4) \right] \, , 
\label{(3.13)}
\end{equation}
where we use the standard definitions $\bar\alpha = \alpha_s N_c / \pi$
and the Lipatov function
\begin{equation}
\chi (\lambda ) = \psi (1) - \frac{1}{2} \psi (\lambda ) - \frac{1}{2} 
\psi ( 1 - \lambda) , \,\, \, \, 
\psi (\lambda ) = \frac{\Gamma^\prime (\lambda)}{\Gamma (\lambda )} \,  . 
\label{(3.14)}
\end{equation}
As usual the integration contour being parallel to the imaginary axis with 
$0 < Re (\lambda ) < 1$. Since we are in the following mainly 
interested in the region in which $q_\bot (\simeq 1/x_\bot )$ is not 
very much larger than the saturation scale $\bar Q_s$, we keep in 
(\ref{(3.13)}) the scale $\bar Q_s (\vec b)$ independent on $\vec x_\bot$.
 The normalization of (\ref{(3.13)}) at $Y=0$ is given by the expression 
(\ref{(3.10)}). This is best seen from the inverse Mellin transform 
in terms of the relation $(t \hat= x^2_\bot )$, 
\begin{equation}
\int^\infty_0 dt \, t^{\lambda-2} (1-e^{-t}) = - \Gamma (\lambda - 1) , 
\label{(3.15)}
\end{equation}
confirmed by partial integration. 

From the definition of $\phi_G$ (\ref{(III.1)}) and using 
\begin{equation}
\int d^2 x_\bot e^{i \vec q_\bot \cdot \vec x_\bot} (x^2_\bot )^{-\lambda}
 = \pi \frac{\Gamma (1 - \lambda )}{\Gamma (\lambda ) } \left( \frac{
q^2_\bot}{4} \right)^{\lambda - 1} , 
\label{(3.16)}
\end{equation}
actually valid for $1/4 < Re (\lambda ) < 1$, we obtain the 
Mellin representation of the unintegrated gluon function, 
\begin{equation}
\phi_G (\vec b , \vec q_\bot , Y) = \frac{N_c}{2\pi^2 \alpha_s} \int_C 
\frac{d\lambda}{2 \pi i} \Gamma (2 - \lambda ) \exp \left[ 2\bar\alpha
\chi (\lambda ) Y - (1 - \lambda ) \ln \left( \frac{q^2_\bot}{\bar Q^{~2}_s 
(\vec b )} \right) \right] . 
\label{(3.17)}
\end{equation}
As a consistency check one obtains the result (\ref{(3.11)}) at $Y=0$
by keeping only the pole at $\lambda=2$, dominating the behaviour at small 
values of $q^2_\bot$. Summing the contributions of all poles, $\lambda = 2, 
3, ...$, one obtains 
\begin{equation}
\phi^{MV}_G (\vec b , \vec q_\bot , Y=0) = \frac{N_c}{2 \pi^2 \alpha_s} 
\frac{q^2_\bot}{\bar Q^2_s (\vec b)} \exp \left[ - \frac{q^2_\bot}{
\bar Q^{~2}_s (\vec b) } \right] , 
\label{(3.18)}
\end{equation}
in case of a ``frozen'' scale $\bar Q^{~2}_s$. Inserting (\ref{(3.18)})
into (\ref{(III.2)}) gives back (\ref{(3.10)}). 

Up to the normalizing factors the function $\phi_G$ in (\ref{(3.17)})
has the structure of the amplitude $T ( Q , \mu , Y)$ discussed in \cite{MT,T},
when identifying $Q = q_\bot$ and $\mu = \bar Q_s (\vec b)$.

Following the same steps as described in the paper \cite{MT},
we consider the solution of $\phi_G$ for large values of $\alpha_s Y$ 
extended into the geometric scaling region \cite{IIM}.
This is achieved   by demanding that $\phi_G (\vec b , \vec q_\bot , Y)$
vanishes close to the saturation boundary, i.e. for 
$q^2_\bot < Q^2_s (\vec b, Y)$, to be defined below. This pragmatic 
procedure includes non-linear effects which are present in the 
Balitsky-Kovchegov equation for $N (\vec b, \vec x_\bot , Y)$ \cite{BK}. 
This characteristic behaviour is also discussed in \cite{Munier}
from a more mathematical point of view.
 It is achieved by a linear superposition
of two BFKL type solutions by shifting the positions of their maxima by a 
finite amount. The final scaling solution, expressed in terms of the $Y$
dependent saturation momentum 
\begin{equation}
Q^2_s (\vec b, Y) = c_s \bar Q^{~2}_s (\vec b) \, \frac{ \exp \left[ 2 
\bar\alpha \, \frac{\chi (\lambda_0)}{1-\lambda_0} Y \right]}{(\alpha_s Y)
^{\frac{3}{2 (1-\lambda_0)}}} \,  , 
\label{(3.21)}
\end{equation}
is (for the case of constant $\alpha_s$), 
\begin{eqnarray}
\phi_G (\vec b , \vec q_\bot , Y) &=&  \phi^{{\rm max}}_G \, (1 - \lambda_0 ) 
\exp \left[ - (1 - \lambda_0 ) \ln \frac{q^2_\bot}{Q^2_s (\vec b , Y)} \right]
\nonumber \\
& \times  & \left[ \ln \left( \frac{q^2_\bot}{Q^2_s (\vec b , Y)} \right) + 
\frac{1}{1-\lambda_0} \right] , 
\label{(3.22)}
\end{eqnarray}
where  $c_s$ and $\phi^{{\rm max}}_G = O(1/\alpha_s)$ are constants.
The value of the  anomalous dimension  $\lambda_0$ is determined by 
\begin{equation}
\frac{\chi^\prime (\lambda_0)}{\chi (\lambda_0)} = - \frac{1}{1-\lambda_0}, 
\,\,  \, \lambda_0 = 0.372 \, .
\label{(3.19)}
\end{equation}
 Obviously $\phi_G$ is maximal,
$\phi_G = \phi^{{\rm max}}_G$, when $q_\bot = Q_s (\vec b, Y)$,
and $\phi_G = 0$ for 
$q_\bot \le Q_s (\vec b, Y) \exp{(-\frac{1}{2 (1-\lambda_0)})}$.

It is well known that this leading order calculation with fixed coupling
yields a large exponent in (\ref{(3.21)}), namely
$2 \bar\alpha \, \frac{\chi (\lambda_0)}{1-\lambda_0} = 4.66.. \alpha_s$,
which is too large to agree with phenomenology \cite{IIMu,GBW}.
However, this discrepancy is resolved in \cite{T},
using the next-to-leading BFKL formalism, which as a result
reduces the exponent to a value in agreement with the Golec-Biernat and
W\"usthoff model \cite{GBW}.

It is important to note that this analytical function (\ref{(3.22)})
successfully compares with the numerical studies \cite{KW,geom}
of the Kovchegov equation. Indeed in \cite{KW}
a good fit by (\ref{(3.22)}) is obtained for a fixed value of the 
anomalous dimension, $\lambda_0 = 0.37$, and for
 $5 < q_\bot / Q_s (Y) < 1000$, 
mainly because of  the logarithmic factor, $\ln \left( \frac{
q^2_\bot}{Q^2_s (\vec b, Y)} \right)$, 
which is present in (\ref{(3.22)}). This comparison also  indicates that the
scaling behaviour is rather rapidly approached.

The $A$, respectively the number of participants
$N_{part}$, dependence of the unintegrated gluon distribution 
(\ref{(3.22)})  is dominated by the behaviour for large $A$ by 
\begin{equation}
\phi_G (\vec b = 0 , \vec q_\bot , Y) \simeq A^{ \frac{1-\lambda_0}{3}}
 \ln A  \, , 
\label{(3.23)}
\end{equation}
rather than by $A^{1/3}$. This is leading twist gluon shadowing due to the 
anomalous behaviour of $\phi_G$,
with a non-vanishing value of $\lambda_0$,
 in the extended  geometrical scaling window
\begin{equation}
Q^2_s (\vec b , Y) \le q^2_\bot < Q^2_s (\vec b , Y) 
\exp{\sqrt{4 \bar\alpha \chi^{\prime \prime} (\lambda_0) Y}} \, ,
\label{(3.24)}
\end{equation}
and because of  $Q^2_s (\vec b = 0 , Y) \simeq \rho R_A \simeq A^{1/3}$.
We note that the upper bound given in (\ref{(3.24)}),
which has its origin in the diffusion region, is more stringent with respect
 to its $Y$ dependence than the one quoted 
in \cite{IIM},
which behaves as $Q_s^4 (\vec b, Y)/{\bar Q_s^2(\vec b)}
 \approx
 \exp{[4 \bar \alpha_s \frac{\chi(\lambda_0)}{1-\lambda_0} Y]}$.


Also when compared to the LO perturbative behaviour, $\phi^{\rm{LO}}_G 
\simeq A$ given by (\ref{(B6)}), even stronger 
suppression of the gluon density  (\ref{(3.23)}) is observed. 

The consequences of these derived scaling properties of $\phi_G$ in
(\ref{(3.22)}) for the 
dilepton cross section are analyzed in the next Section.

\section{Anomalous scaling and shadowing in dilepton production}

\subsection{Qualitative results}

Before analyzing the transverse $\gamma^*$ differential cross section
in the $k_\bot$-factorized form of
(\ref{(11)}) in more detail we first investigate its scaling properties.
 We define
\begin{equation}
\frac{d\sigma^{qA \rightarrow \gamma^* X}}{d^2 b} \equiv (k^2_\bot + \eta^2) 
\frac{d\sigma^{qA \rightarrow \gamma^* X}}{d^2 b d^2 k_\bot d \ln z} \, . 
\label{(5.1)}
\end{equation}
Assuming $\eta < k_\bot$, such that  $k_\bot$  is the hard scale, we may
approximate this cross section by 
\begin{equation}
\frac{d\sigma^{qA \rightarrow \gamma^* X}}{d^2 b} = \int 
\frac{d^2 q_\bot}{\pi q^2_\bot}
H (\vec k_\bot , z \vec q_\bot , z) \phi_G (\vec b , \vec q_\bot / Q_s 
(\vec b , Y)) \, , 
\label{(5.2)}
\end{equation}
with
\begin{equation}
H (\vec k_\bot , \vec q_\bot , z) = \frac{\alpha_{em} \alpha_s}{N_c} 
[1 + (1-z)^2] \,  \frac{\vec q^{~2}_\bot}{(\vec k_\bot - \vec q_\bot )^2
+ \eta^2} , 
\label{(5.3)}
\end{equation}
(similar to the definition given in the previous section). 

Inserting the scaling function (\ref{(3.22)}), the cross section
(\ref{(5.2)}) scales approximately as follows, 
\begin{equation}
\frac{d\sigma^{qA \rightarrow \gamma^* X}}{d^2 b} = 
\phi_G (k_\bot / (z Q_s (\vec b , Y)), 
\vec b) \,  F \left[ \ln \left( \frac{k_\bot}{z Q_s (\vec b , Y)} \right) , 
\eta / k_\bot , z \right] \, ,
\label{(5.4)}
\end{equation}
where $F$ is expected to be a slowly varying function of $k_\bot$.
In order to exhibit the anomalous $A$ dependence
we deduce a parametric estimate of the ratio with respect to the proton target,
\begin{equation}
R_{pA} = \frac{d\sigma^{q A \rightarrow \gamma^* X}/d^2 b}
{\rho T(b) \sigma^{q p \rightarrow \gamma^* X}} \, .
\label{(5.5)}
\end{equation}
For central collisions, $\vec b = 0$, and 
assuming that the  extended geometric scaling regions for protons $p$ and
nuclei $A$ indeed overlap,
this ratio becomes
\begin{equation}
R_{pA} \approx 
A^{-\lambda_0 / 3} \, .
\label{(5.5a)}
\end{equation}
Because of the nonvanishing anomalous dimension $\lambda_0$,
 we thus predict shadowing of  $\gamma^*$ production
in quark-nucleus scattering  at fixed $k_\bot$ and $Y$, at a constant level. 
The estimate (\ref{(5.5a)})
 is based on approximating (\ref{(3.22)}) by 
\begin{equation}
\phi_G ( k_\bot / Q_s (\vec b, Y)) \approx \left( k^2_\bot / Q^2_s (\vec b, Y)
\right)^{\lambda_0 - 1} , 
\label{(5.7)}
\end{equation}
and on (c.f. (\ref{(3.7)}) and (\ref{(3.21)}) ) 
\begin{equation}
{Q^2_s (\vec b  , Y) \Big|_A} \approx \rho T(b) \, ,
\label{(5.8)}
\end{equation}
whereas the scale $Q^2_s ( Y) \Big|_p$
of the proton does not depend on $\vec b$.

A similar suppression in terms  of anomalous scaling,
 as given e.g.  by (\ref{(5.5a)}),
is also predicted for the nuclear modification factor 
$R^{G}_{pA}$ in case of
gluon production \cite{KLM,bkw,KW,cronin,JNV,JJM:2004,Iancu:2004}.

\subsection{Quantitative results}

For illustration we present numerical estimates for the transverse
$\gamma^*$ differential cross section (\ref{(11)}),
 for the
RHIC energy $\sqrt{s} = 200$ GeV.
We are estimating the ratio $R_{pA}$  (\ref{(5.5)}) as the ratio of central,
$\vec b = 0$, versus peripheral, $\vec b > 0$,
 $qA \rightarrow \gamma^* X$ collisions,
as follows
\begin{equation}
R_{pA} = \frac{d\sigma^{q A \rightarrow \gamma^* X}/d^2 b}
{\rho T(b)}\Big|_{central} \, {\Big /} \,
  \frac{d\sigma^{q A \rightarrow \gamma^* X}/d^2 b}
{\rho T(b)}\Big|_{peripheral} \,  \, ,
\label{(5.55)}
\end{equation}
where we choose, according to (\ref{(5.8)}),
\begin{equation}
\frac{\rho T(b)\Big|_{peripheral}}{\rho T(b)\Big|_{central}}
 = \frac{Q^2_s (\vec b, Y)\Big|_{peripheral}}{Q^2_s (\vec b =0, Y)} \, .
\label{(5.8a)}
\end{equation}
The peripheral collision is assumed to be such, that
$N_{part} = 1$, i.e. the proton.  For the numerics the ratio (\ref{(5.8a)})
is taken
to be equal to  $\Bigl( Q_{s,p} {\Big /} Q^{MV}_{s,A} \Bigr)^2 
\simeq A^{-1/3}$.  
The central (peripheral) $\gamma^*$ cross section in (\ref{(5.55)})
 is evaluated 
at the scale $Q_s (\vec b =0, Y) ~~$
($Q_s (\vec b, Y)\vert_{peripheral} = A^{-1/6}
Q_s (\vec b =0, Y)$).

\begin{figure}[htb]
\begin{center}
\epsfig{bbllx=0,bblly=0,bburx=600,bbury=465,
file=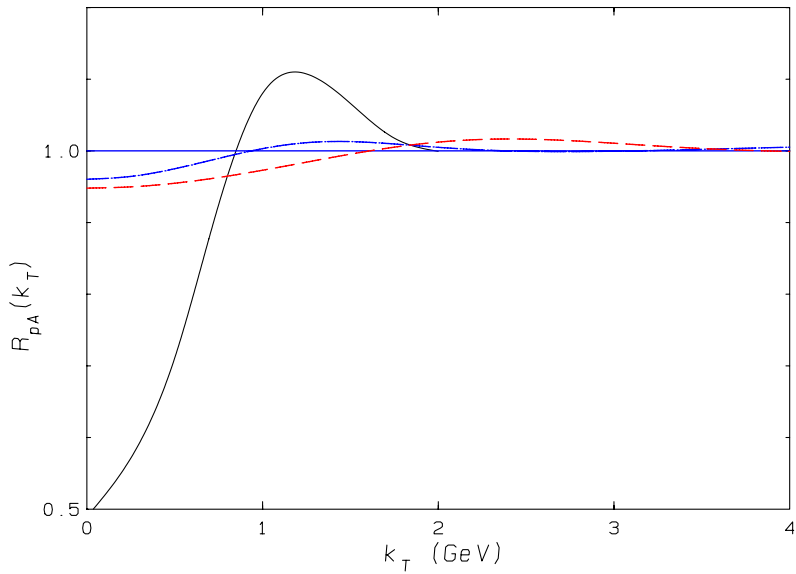,width=10.5cm}
\caption{McLerran-Venugopalan model: 
$R_{pA}$ as a function of $k_\bot$. Solid curve for
$ M = 2 $ GeV and fixed $z = 0.7$; dashed (dot-dashed) curve for$ M = 4 (2)$
 GeV and $y_{\gamma} = 3$.}
\label{fig4}
\end{center}
\end{figure}


 In order to set the reference we
give results based on the McLerran-Venugopalan model \cite{MV} as described
in Sec.~3, when using $\phi_G^{MV}$ of  (\ref{(3.18)}).
Instead of explicitly taking the scale (\ref{(3.7)}) we fix the values
at $\vec b = 0$
by $Q^{MV}_{s,A} = 1$ GeV for $A = 200$, and $Q_{s,p} = 1/A^{1/6}$ GeV 
$\simeq 0.41$ GeV for the proton target, respectively.

\begin{figure}[htb]
\begin{center}
\epsfig{bbllx=0,bblly=0,bburx=600,bbury=460,
file=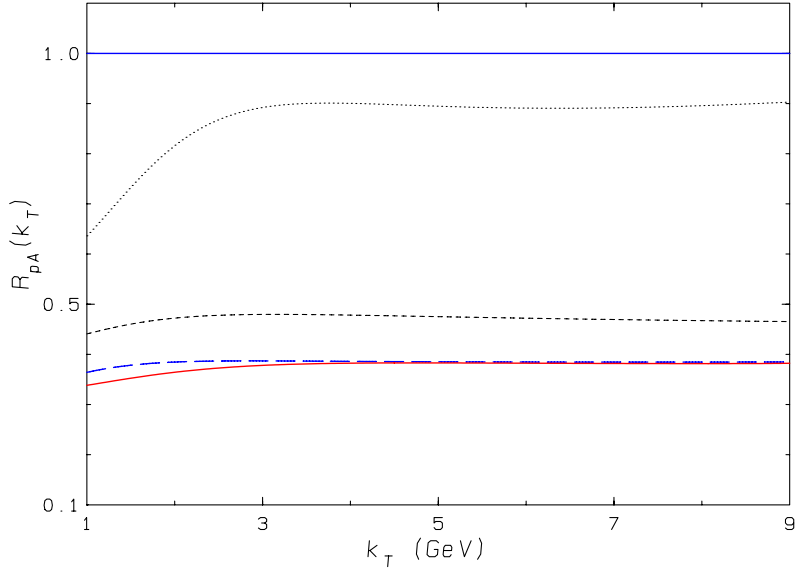,width=10.5cm}
\caption{BFKL-saturation model: 
$R_{pA}$ as a function of $k_\bot$ for different values of   $y_{\gamma}$
and for $M = 2$ GeV: dotted ($y_{\gamma} = 0.5$), short-dashed ($y_{\gamma} =
1.5$), long-dashed ($y_{\gamma} = 3.0$).
 Solid curve for
$ M = 4 $ GeV and  $y_{\gamma} = 3$.}
\label{fig5}
\end{center}
\end{figure}


\begin{figure}[htb]
\begin{center}
\epsfig{bbllx=0,bblly=0,bburx=600,bbury=460,
file=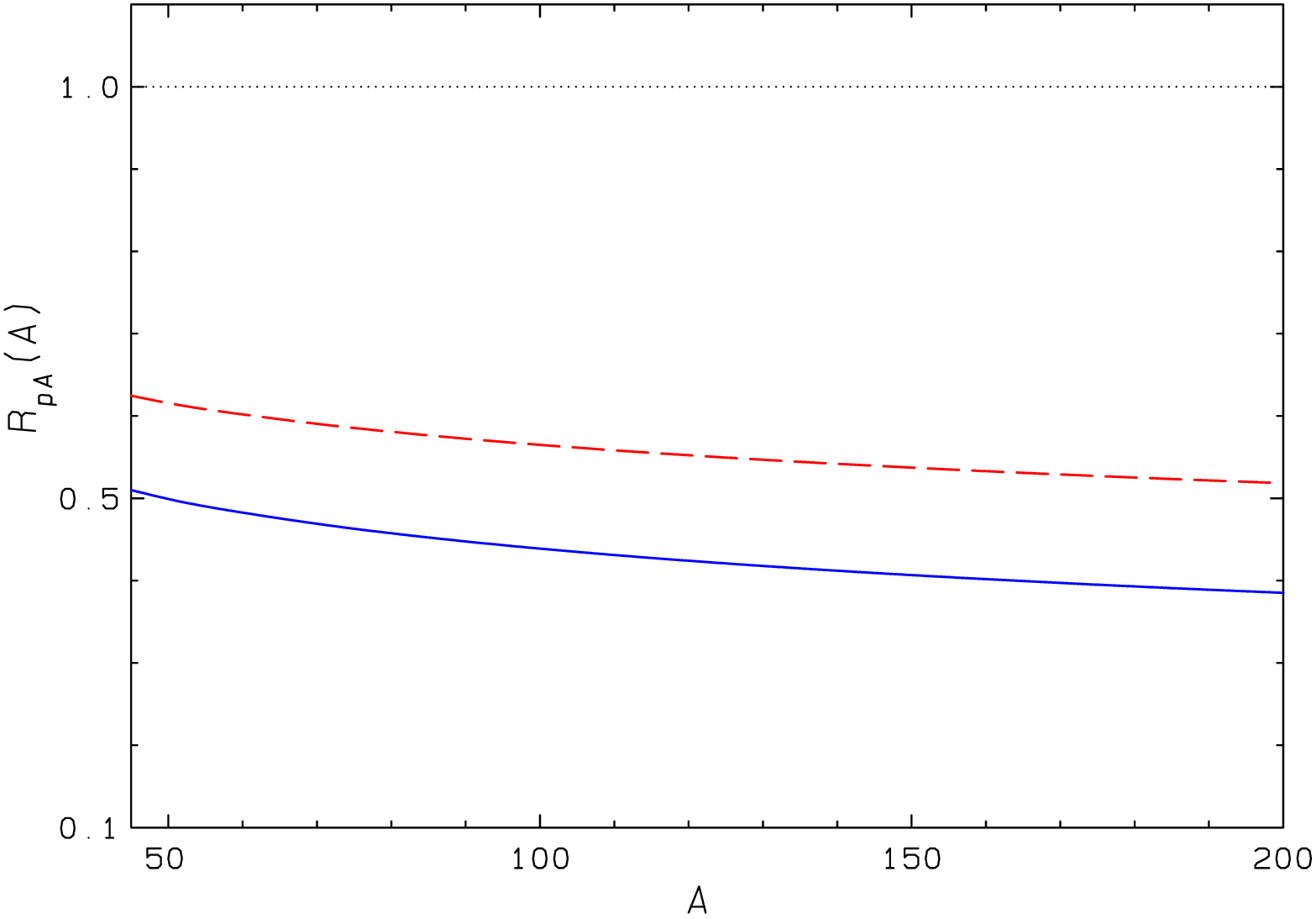,width=10.5cm}
\caption{BFKL-saturation model: 
$R_{pA}$ as a function of $A$ for $k_\bot = 5$ GeV, $y_{\gamma} = 3$
and  $M = 2$ GeV (solid curve). Dashed curve corresponds 
to $A^{-{\lambda_0/3}}$.}
\label{fig6}
\end{center}
\end{figure}


For small photon rapidities, $y_{\gamma} \simeq 0$, and dilepton masses of $M =
2 - 4 $ GeV the ratio $R_{pA}$ of (\ref{(5.55)}) as a function of $k_\bot$
is essentially $R_{pA} = 1$. This is easy to see from (\ref{(11)}):
small values of $z \le 0.05$ imply that $H$ of (\ref{(5.2)}) is
approximated by $H \approx q^2_{\bot}/(k^2_{\bot} + \eta^2)$.
Because of (\ref{(3.12)}) it follows that 
$R_{pA} \approx A^{-1/3} (Q^{MV}_{s,A}/Q_{s,p})^2 \simeq 1$.
For comparison with the prediction for evolved gluons we also consider 
 larger values of $y_{\gamma}$, e.g. $y_{\gamma} = 3$, in this model.
The results for $R_{pA}$ are plotted in Fig.~4 for two values of
$M=2$ and $M=4$ GeV, respectively. A very small suppression,
$R_{pA} \ge 0.95$, is observed for $k_\bot \le 1 - 2$ GeV.
Keeping, however, the value of $z$ fixed and large, e.g. $z = 0.7$,
a significant Cronin type pattern for the transverse Drell-Yan
spectrum is observed (solid curve in Fig.~4): shadowing 
for $k_\bot \le 1$ GeV, and enhancement above. This confirms the results first
presented in \cite{Kopeliovich}. However, we have to keep in mind
that this model is only reliable, when $k_\bot > Q^{MV}_s$.

In order to obtain results at large photon rapidities based
on the BFKL evolution in the presence of saturation
we pragmatically have to choose the $Y$ dependence of the scale,
instead of the one given by (\ref{(3.21)}). 
As already discussed in Sec.~3 we take the one compatible
with phenomenology, following \cite{GBW},
 \begin{equation}
Q^2_s (\vec b = 0 , Y =  y_{\gamma}) =
 (Q^{MV}_s)^2 \exp(\lambda_{GBW} \,  y_{\gamma}) , \, \,
 \lambda_{GBW} = 0.3  \, \, .
\label{(5.9)}
\end{equation}

In accordance with the discussion in  Sec.~5.1 significant shadowing
is obtained, as shown in Fig.~5 as a function of $k_\bot$,
especially when the dileptons are produced rather forward, e.g. 
with  $y_{\gamma} = 3$.
Similar results, however, for 
$k_\bot$-integrated dilepton rates are presented in \cite{JMJ}.

When $k_\bot >> Q_s (\vec b = 0 , Y)$ the ratio $R_{pA}$
becomes essentially independent on the transverse momentum.

Finally, we numerically check the statement in (\ref{(5.5a)})
concerning the $A$ dependence of the $\gamma^*$ cross section ratio
for central collsions.
An example is plotted in Fig.~6, indicating that indeed the anomalous
$A$ dependence is  to be expected for large nuclei, with the consequence
of strong (leading twist) shadowing of photons/dileptons,
 when produced in the forward direction
of $pA$,  or $dA$ collisions.

\subsubsection*{Acknowledgments}

The authors gratefully acknowledge  helpful discussions  
with  Francois Gelis and Jamal Jalilian-Marian.
R.~B. is supported, in part, by DFG, contract FOR 329/2-1, and
A.~M. is supported, in part, by the US Department of Energy.

\def\appendix{\par
 \setcounter{section}{0}
 \setcounter{subsection}{0}
 \def\thesection{Appendix \Alph{section}}
 \def\thesubsection{\Alph{section}.\arabic{subsection}}
 \def\theequation{\Alph{section}.\arabic{equation}}
 \setcounter{equation}{0}}

\appendix

\section{}

\setcounter{equation}{0}

In this appendix we relate the $k_\bot$-representation of the $\gamma^*$ 
production cross section, Eq.(\ref{(11)}), 
to its impact parameter representation \cite{Brodsky, Kopeliovich, 
Raufeisen}.
In order to go from $\vec k_\bot$ to the conjugate coordinate $\vec x_\bot$,
 we introduce the  Fourier transform of
the propagator \cite{Mueller:1999wm},
\begin{eqnarray}
\frac{\vec\epsilon^{~\lambda}_\bot \cdot \vec k_\bot}{k_\bot^2 + \eta^2} & = & 
-i \int \frac{d^2 x_\bot}{2\pi} \left( \vec\epsilon^{~\lambda}_\bot \cdot
 \vec\nabla_
{x_\bot} e^{i\vec k_\bot \cdot \vec x_\bot} \right) K_0 (\eta x_\bot)
\label{(A1)} \\
& = & - i \eta \int \frac{d^2 x_\bot}{2\pi}
 e^{i \vec k_\bot \cdot \vec x_\bot} 
\frac{\vec\epsilon^{~\lambda}_\bot \cdot \vec x_\bot}{x_\bot}
 K_1 (\eta x_\bot), 
\label{(A2)}
\end{eqnarray}
with $x_\bot = | \vec x_\bot |$ and $K_i$ the Bessel functions  of the second
kind, $K_1 (z) = - \frac{d}{dz} K_0 (z)$.

 The  square of the radiation 
amplitude (\ref{(7)}) 
is expressed by 
\begin{eqnarray}
\sum_{\lambda} & &  \left|
 \frac{\vec\epsilon^{~\lambda}_\bot \cdot \vec k_\bot}
{k_\bot^{~2} 
+ \eta^2} - 
\frac{\vec\epsilon^{~\lambda}_\bot \cdot \vec k_\bot^{~\prime}}{k_\bot^
{~\prime 2} + \eta^2} \right|^2  =  \nonumber \\
& & \int \frac{d^2 x_\bot d^2 y_\bot}{(2\pi)^2} e^{i \vec k_\bot \cdot 
(\vec x_\bot - \vec y_\bot)} 
\, \eta^2 \frac{\vec x_\bot \cdot \vec y_\bot}{x_\bot y_\bot} K_1 
(\eta x_\bot ) K_1 (\eta y_\bot ) \label{(A3)} \\
& & \times \left[ \left( 1 - e^{-iz \vec q_\bot \cdot x_\bot} \right) + 
\left( 1 - e^{+ i z \vec q_\bot \cdot \vec y_\bot} \right) - \left( 1 - 
e^{-iz \vec q_\bot \cdot ( \vec x_\bot - \vec y_\bot)}\right) \right]
\nonumber
\end{eqnarray}
Following \cite{Kopeliovich}, we
introduce the transverse photon
 wave function for  $q \rightarrow \gamma^* q$,
$~\psi^\lambda_{\gamma q} (z , x_\bot , \eta)$, with
\begin{eqnarray}
& & \sum_{\lambda}  \psi^{\lambda^*}_{\gamma q} (z , \vec x_\bot , \eta )
\psi^\lambda_{\gamma q} (z , \vec y_\bot , \eta ) = \nonumber \\
& & \frac{\alpha_{em}}{2\pi^2} 
[ 1 + ( 1 - z)^2 ] 
 \,  \eta^2 \, \frac{\vec x_\bot \cdot \vec y_\bot}{x_\bot y_\bot} K_1 (\eta
x_\bot) K_1(\eta y_\bot )  \, . 
\label{(A4)}
\end{eqnarray}
Inserting (\ref{(III.2)}) for the unintegrated gluon function $\phi_G$,
namely 
\begin{equation}
\int \frac{d^2 q_\bot}{q^2_{\bot}}
 \phi_G ( \vec b, \vec q_{\bot} , Y) \left( 1 - e^{-iz \vec q_\bot \cdot 
\vec x_\bot} \right) = \frac{N_c}{2 \pi \alpha_s} N (\vec b, z \vec x_\bot , Y)
\, ,
\label{(A5)}
\end{equation} 
into
the $\gamma^*$ cross section  (\ref{(11)})
we finally obtain,
together with (\ref{(A3)}) the impact parameter representation
 of the $\gamma^*$ cross section \cite{Kopeliovich},
\begin{eqnarray}
& &\frac{d\sigma^{qA \rightarrow \gamma^* X}}{d^2 b d^2 k_\bot d \ln z} 
= \int \frac{d^2 x_\bot d^2 y_\bot}{(2 \pi)^2}
e^{i \vec k_\bot \cdot(\vec x_\bot - \vec y_\bot)} \label{(A8)} \\
& & \times \sum_{\lambda} \psi^{\lambda^*}_{\gamma q}
 (z , \vec x_\bot , \eta ) \psi
^\lambda_{\gamma q} (z , \vec y_\bot , \eta ) \left[ 
N (\vec b , z \vec x_\bot , Y) + N ( \vec b , z \vec y_\bot , Y)
- N (\vec b , z ( \vec x_\bot - \vec y_\bot ) , Y) \right]  . \nonumber 
\end{eqnarray}
In the $q \bar q$-dipole approximation for scattering off a nucleus $A$, the 
amplitude $N = 1 - S$, where
\begin{equation}
S ( \vec b , \vec x_\bot , Y) = 
\exp [ - \frac{1}{2} \sigma_{q \bar q} (\vec b , 
\vec x_\bot , Y) ] = \exp [ -  x^{~2}_\bot \bar Q^{~2}_s (\vec b) / 4 ] \, , 
\label{(A9)}
\end{equation}
is expressed by the $  \sigma_{q \bar q}$ cross section or by
the saturation scale 
$\bar Q^{~2}_s (\vec b ) = \frac{C_F}{N_c} Q^2_s (\vec b )$
given in (\ref{(3.7)})\cite{Mueller:2001fv,Mueller:1999wm}. 

Let us add two useful relations:
\begin{itemize}
\item[i)]
The limiting case for real photon emission, $\eta \rightarrow 0$, is obtained 
from (\ref{(A4)}) by 
\begin{equation}
\sum_{\lambda} \psi^{\lambda^*}_{\gamma q} (z , \vec x_\bot ) \psi^\lambda_
{\gamma q} (z , \vec y_\bot ) \rightarrow \frac{\alpha_{em}}{2\pi^2} 
[1 + (1-z)^2] \frac{\vec x_\bot \cdot \vec y_\bot}{x^2_\bot y^2_\bot} \, .
\label{(A11)}
\end{equation}
\item[ii)]
Via crossing from $q \rightarrow q \gamma^* (M)$ into $\gamma^* (Q) 
\rightarrow q \bar q$ the DIS cross section is obtained from the expression
(\ref{(A8)}) by the substitutions 
\begin{equation}
z \rightarrow \frac{1}{1 - \alpha },  \, \,{\rm i.e.}  \, \, 
\frac{1 + (1 - z)^2}{z^2}  \rightarrow (1 - \alpha )^2 + \alpha^2 , 
\label{(A12)}
\end{equation}
and 
\begin{equation}
K_1 (\eta x_\bot ) = K_1\left( \sqrt{(1 - z) M^2 x^2_\bot} \right) \rightarrow
K_1 \left( \sqrt{ \alpha (1 - \alpha ) Q^2 {\bar x}^2_\bot} \right) , 
\label{(A13)}
\end{equation}
where the $q \bar q$ separation is given by
$\bar x_\bot = \frac{x_\bot}{1 - \alpha} = z x_\bot$, such that 
\begin{equation}
N ( \vec b , z \vec x_\bot , Y)
 \rightarrow N ( \vec b , \vec{\bar x}_\bot , Y)\, .
\label{(A14)}
\end{equation}
The explicit expressions for deep inelastic scattering may be found 
in \cite{Brodsky,Mueller:2001fv}.

\end{itemize}

\section{}
\setcounter{equation}{0}

Here we give some details on the behaviour of the leading order (LO)
cross section for real and isolated photons produced via the process 
$q A \rightarrow \gamma X$
\cite{Gelis:2002ki}. The photon takes away a large transverse momentum
$k_\bot , k_\bot \gg Q_s$, opposite to a recoil quark jet. 

The differential cross section for real photons, $M^2 = 0$, in the 
$k_\bot$-factorized form reads (see Eq.~(\ref{(11)})), 
\begin{equation}
\frac{d\sigma}{d^2 b d^2 k_\bot d \ln z} = \frac{\alpha_{em} \alpha_s}{
\pi N_c} \frac{1 + (1-z)^2}{k^2_\bot} \int d^2 q_\bot \frac{\phi_G 
(\vec b, \vec q_\bot , Y)}{[\vec q_\bot - \vec k_\bot / z]^2} .
\label{(B1)}
\end{equation}
Assuming $k_\bot \gg z \vec q_\bot$, i.e. such that collinear quark-isolated 
photon configurations are suppressed,
we may  write (\ref{(B1)}) as 
\begin{equation}
\frac{d\sigma}{d^2 k_\bot d \ln z} = \frac{\alpha_{em} \alpha_s}{N_c}
\frac{ z^2 [ 1 + (1 - z)^2]}{k^4_\bot} x G_A \left( x = \frac{x^2_T}
{z (1-z)} , Q^2 = k^2_\bot \right) ,
\label{(B2)}
\end{equation}
where $x G_A$ is the gluon distribution in the nucleus $A$, 
\begin{equation}
x G_A (x , k^2_\bot) = \int^{O(k_\bot)}
 \frac{d^2 q_\bot}{\pi} \int d^2 b \phi_G 
( \vec b , \vec q_\bot , Y) . 
\label{(B3)}
\end{equation}
in agreement with (\ref{(4)}).
For real photons $x_T = k_\bot /\sqrt{s}$ and $z = x_\bot \exp{y_{\gamma}}$,
such that the $x$ value in the gluon is decreasing with increasing
photon rapidity $y_{\gamma}$.
The expression (\ref{(B2)})
is the same as derived from the hard LO pQCD $2\rightarrow 2$ Compton process 
$q G \rightarrow \gamma q$ for production of isolated photons at large 
$k_\bot$ \cite{Ellis}: indeed the corresponding hard cross section is given 
by 
\begin{eqnarray}
\frac{d \sigma}{d \hat t}^{qG \rightarrow \gamma q} &=& \frac{\pi \alpha_{em}
\alpha_s}{N_c} \, \frac{1}{\hat s^{~2}} \left[ - \frac{\hat s}{\hat t} - 
\frac{\hat t}{\hat s} \right] \nonumber \\
& = & \frac{\pi \alpha_{em} \alpha_s}{N_c} \, \frac{z^2 (1-z) [1 + 
(1-z)^2]}{k_\bot^4} , 
\label{(B3a)}
\end{eqnarray}
since $\hat s = \frac{k^2_\bot}{z(1-z)}, \, \hat t = - k^2_\bot / z$.
 After folding (\ref{(B3a)}) into the expression for 
$qA \rightarrow \gamma X$ with the help of the gluon function $xG_A$ in the 
nucleus $A$ and performing the integration over the recoiling quark 
leads to  (\ref{(B2)}).   

Using (\ref{(III.1)}), we relate the gluon distribution $\phi_G$
to the dipole amplitude $N_{q \bar q}$
in the quasi-classical approximation \cite{MV}.
With (\ref{(A9)}) 
one obtains, 
\begin{equation}
\phi_G (\vec b , \vec q_\bot , Y) = \frac{N_c}{(2\pi)^3 \alpha_s} \int
d^2 x_\bot e^{i \vec q_\bot \cdot \vec x_\bot} \vec\nabla^2_{x_{\bot}} 
[ 1 - \exp ( - x^2_\bot \bar Q^2_s / 4)],
\label{(B5)}
\end{equation}
with the scale given by (\ref{(3.7)}).
At very large $q_\bot \gg Q_s$ one finds, 
\begin{equation}
\phi_G^{\rm{LO}} (\vec b , \vec q_\bot , Y) =
\rho T(b) 
\, \frac{\alpha_s  N_c C_F}{q^{~2}_\bot} \, . 
\label{(3.5)}
\end{equation}
This is derived by using the LO gluon distribution,
\begin{equation}
x G^{{\rm LO}} (x , 1/x^2_\bot) = \frac{C_F N_c 
\alpha_s}{\pi}\ln \frac{1}{x^2_\bot\Lambda^2} , 
\label{(BB6)}
\end{equation}
and thus
\begin{equation}
N^{\rm{LO}} (\vec b , \vec x_\bot , Y)=  x^2_\bot \bar Q^{~2}_s (\vec b)/4
= \frac{C_F \alpha^2_s}{2} \rho T(b) 
x^2_\bot \ln \frac{1}{x^2_\bot \Lambda^2}  \, .
\label{(3.6)}
\end{equation}
 (\ref{(3.5)}) follows
by using \cite{Kovchegov:2000hz}
\begin{equation}
\int \frac{d^2 x_\bot}{(2\pi)^2} e^{i\vec q_\bot \cdot \vec x_\bot}
\ln \left( \frac{1}{x^2_\bot \Lambda^2} \right) = \frac{1}{\pi q^2_\bot} ,
\label{(3.9)}
\end{equation}
valid for $\vec q_\bot \not= 0$.
Since
\begin{equation}
\int \phi^{{\rm LO}}_G (\vec b , \vec q_\bot , Y) d^2 b =
 A \frac{\alpha_s C_F N_c}{\pi} \frac{1}{q^2_\bot} , 
\label{(B6)}
\end{equation}
the LO factorized result for the gluon in  nucleus $A$,
\begin{equation}
x G^{{\rm LO}}_A (x , k^2_\bot) \simeq A \frac{(N^2_c - 1) \alpha_s}{2\pi} \ln 
\frac{k^2_\bot}{\Lambda^2} = A x G^{{\rm LO}} (x , k^2_\bot ) , 
\label{(B7)}
\end{equation}
in terms of the ${\rm LO}$
gluon in the nucleon, is finally obtained.


\section{}
\setcounter{equation}{0}

In this appendix we extensively discuss in more general terms  
the relationship between the 
$k_\bot$-factorization, which has been used 
in the previous sections and the usual QCD factorization 
\cite{Ellis} involving local gauge 
invariant operators which appear in a Wilson operator product expansion.
As in Appendix B, where we investigate this relationship for large
$k_\bot >> Q_s$ in LO pQCD,
 we concentrate explicitly on the case of
real photon production.

\subsection{The forms of factorization}

 We  restrict here our discussion to the scaling region where the 
hard scale is above, but not too far above, the saturation region.
 When the hard
scale is below the saturation momentum neither $k_\bot$-factorization nor the 
usual QCD factorization 
is applicable as higher twist terms are coherent with 
leading twist terms and all terms must be considered together. While it 
does make sense to talk of a gluon distribution which has
 reached saturation, that
distribution does not appear simply in factorization formulae.
When the hard scale is very large and outside the scaling region 
$k_\bot$-factorization may still be useful, but the issues are more 
straightforward than in the scaling region. 

We write generically a dimensionless observable in the 
$k_\bot$-factorized form 
\begin{equation}
\sigma (\vec Q, Y) = \int \frac{d^2 q_\bot}{\pi q^2_{\bot}} \,
H (\vec q_{\bot} , \vec Q ) \, 
\phi_G (\vec  q_{\bot} , Y) \, ,
\label{(4.5)}
\end{equation}
and in the QCD factorization form 
\begin{equation}
\sigma (\vec Q , Y) = \frac{1}{Q^2} \int^\infty_0 dy \,  \tilde H (y) \,
x G ( \vec Q , Y-y) \, ,
\label{(4.6)}
\end{equation}
where $Q$ is the hard scale of the reaction,
 $H$ is the hard part in the $k_\bot$-factorized form, and 
$\tilde H$ the hard part in the usual factorization. 
Here, and in the following, we suppress writting explicitly 
the dependence on  the impact parameter  $\vec b_{\bot}$.
We suppose that the hard part in the $k_\bot$-factorized expression is not 
too nonlocal in rapidity while we cannot suppose
 such is the case for $\tilde H$. 
Finally the coupling in $H$ and $\tilde H$ should be taken at the hard scale
$Q$. In our example of direct photon production the hard scale $Q$ 
becomes the transverse momentum, $k_\bot$, of the photon, while 
\begin{equation}
\sigma ( \vec k_{\bot} , Y) \equiv
\sigma (\vec b_{\bot} , \vec k_{\bot} , Y) = 
\frac{k^{~2}_\bot d \sigma}{d^2 b_\bot d^2 k_\bot d
  \ln z}  \, , 
\label{(4.7)}
\end{equation}
and (c.f. (\ref{(B1)})) 
\begin{equation}
H (\vec q_{\bot} , \vec k_{\bot} ) = 
\frac{\alpha_{em} \alpha_s}{N_c}\, [1 + (1 - z )^2 ] 
\frac{q^{~2}_\bot}{(\vec q_{\bot} - \vec k_{\bot} / z)^2} \,  .
\label{(4.8)}
\end{equation}
In the scaling region  we approximate $\phi_G$ and $xG$,
respectively,
 from (\ref{(3.22)})
by neglecting in the following possible constants under the logarithms.
We  write it in the form
\begin{equation}
\phi_G (\vec q_{\bot} , Y) = 
\frac{C}{\alpha_s} \left( \frac{Q^2_s (Y)}{
q^{2}_{\bot}} \right)^{1 - \lambda_0} \, \ln (q^2_\bot / Q^2_s (Y))  \, ,
\label{(4.9)}
\end{equation}
and 
\begin{equation}
x G ( Q , Y) = \frac{{\tilde C} Q^2}{\alpha_s} 
\left( \frac{Q^2_s (Y)}{Q^2} \right)^{1-\lambda_0}
 \, \ln ( Q^2 / Q^2_s (Y) )  \, .
\label{(4.10)}
\end{equation}

\noindent
In this region the normalizing factors $C $ and ${\tilde C}$
are not necessarily the same, and actually
we have not been able to relate them.

\subsection{The renormalization group}

In this section we show that (\ref{(4.10)}) obeys the renormalization group
equation. In showing this we shall find a relation which will be crucial in 
relating $H$ and $\tilde H$, which appear in  (\ref{(4.5)}) and (\ref{(4.6)}).
Now in BFKL dynamics there are two alternate forms for $xG$, 
\begin{equation}
xG (Q , Y) = \int \frac{dn}{2 \pi i} A_n \, e^{\gamma_n \ln Q^2 / \mu^2 + 
(n - 1)Y} \, ,
\label{(4.11)}
\end{equation}
and (c.f. (\ref{(3.13)}))
\begin{equation}
xG ( Q , Y) = \int \frac{d\lambda}{2 \pi i} B_\lambda \, e^{2 \bar{\alpha}
 \chi (\lambda ) Y + \lambda \ln Q^2 / \mu^2} \,  ,
\label{(4.12)}
\end{equation}
where the scale $\mu$ is introduced to create dimensionless quantities, but
$xG$ does not depend on $\mu$. The integral in (\ref{(4.11)}) goes parallel to
the imaginary axis with $Re (n)$
 to the right of all singularities of $\gamma_n$
and $A_n$ in $n$. As discussed in Sec.~3, 
Eq.~(\ref{(4.12)}) is, of course,  not 
a perfectly correct representation of BFKL dynamics in the presence of 
saturation, nevertheless the aspects of (\ref{(4.12)}), which we use remain 
true even when the BFKL equation \cite{BFKL} is replaced
 by the Kovchegov equation \cite{BK}.
 
In the scaling region the integrals in (\ref{(4.11)}) and (\ref{(4.12)})
are dominated by saddle points at $n = n_0$ and $\lambda = \lambda_0$, where 
$\lambda_0$ satisfies (\ref{(3.19)}).
 Also for (\ref{(4.11)}) and (\ref{(4.12)}) to describe the 
same function it must be true that 
\begin{equation}
 n-1 = 2 \bar{\alpha} \chi (\gamma_n )\,  , 
\label{(4.13)}
\end{equation}
which determines $\gamma_n$,  while at the saddle points
\begin{equation}
n_0 - 1 =  2 \bar{\alpha} \chi (\lambda_0) \, ,
\label{(4.14)}
\end{equation}
and 
\begin{equation}
\lambda_0 = \gamma_{n_0}  \, .
\label{(4.15)}
\end{equation}

The renormalization group equation \cite{Ellis} is 
\begin{equation}
Q^2 \frac{\partial}{\partial Q^2} x G ( Q , Y) = \int^\infty_0 
dy \, \gamma (y)\,  x G ( Q , Y-y) , 
\label{(4.16)}
\end{equation}
with $\gamma (y)$ the gluon anomalous dimension 
\begin{equation}
\gamma (y) = \int \frac{dn}{2 \pi i} \gamma_n \, e^{(n-1)y} \,  .
\label{(4.17)}
\end{equation}
Using (\ref{(4.10)}) along with the result (\ref{(3.21)}),
\begin{equation}
Q^2_s (y) \simeq  \frac{\exp\left\{\frac{ 2 \bar{\alpha} \chi 
(\lambda_0)}{1 - \lambda_0} \, y \right\}}
{[\alpha_s  y]^{\frac{3}{2(1-\lambda_0)}}}
\, ,
\label{(4.18)}
\end{equation}
it is straight forward to get 
\begin{equation}
xG ( Q , Y-y) = xG ( Q , Y) \, e^{- 2 \bar{\alpha} \chi (\lambda_0)
y} \left[ 1 + \frac{ \frac{ 2 \bar{ \alpha} \chi (\lambda_0)}{ 1-\lambda_0} y}
{\ln (Q^2 / Q^2_s (y))} \right]  \, , 
\label{(4.19)}
\end{equation} 
so long as $y / Y \ll 1$. Using (\ref{(4.10)}) on the left hand side of 
(\ref{(4.16)}), and (\ref{(4.19)}) on the right hand side one easily sees that 
(\ref{(4.16)}) is satisfied if 
\begin{equation}
\lambda_0 = \int^\infty_0 dy \, e^{- 2 \bar{\alpha} \chi (\lambda_0) y}
\, \gamma (y) \, ,
\label{(4.20)}
\end{equation}
and
\begin{equation}
1 = \frac{2 \bar{\alpha} \chi (\lambda_0)}{1-\lambda_0} \int^\infty_0
dy \, y \, \gamma (y) \,  e^{ -  2 \bar{\alpha}  \chi ( \lambda_0) y}
\label{(4.21)}
\end{equation}
are both true.  Eq.(\ref{(4.20)}) follows from (\ref{(4.14)}),
(\ref{(4.15)}) and (\ref{(4.17)}).
 Eq.(\ref{(4.21)}) can be written as 
\begin{equation}
1 = - \frac{2 \bar{\alpha} \chi (\lambda_0)}{1 - \lambda_0} \,
\frac{d}{dn} \gamma_n \Bigg|_{n=n_0} ,
\label{(4.22)}
\end{equation}
which requires the inverse representation of (\ref{(4.17)}).
The validity of (\ref{(4.22)}) then follows from differentiating (\ref{(4.13)})
with respect to $n$ and using (\ref{(3.19)}).

\subsection{The relationship between $k_\bot$-factorization and QCD 
factorization}

We now reach the main topic of this section, namely 
the relationship between the 
two forms of factorization exhibited in (\ref{(4.5)}) and (\ref{(4.6)}).
We begin with (\ref{(4.5)}) and insert (\ref{(4.9)}) for $\phi_G$. Thus 
\begin{equation}
\sigma (\vec Q , Y) = \int \frac{d^2 q_\bot}{\pi q^2_{\bot}} \,
 H (\vec q_{\bot} ,\vec Q) \,
\frac{C}{\alpha_s} 
\left( \frac{Q^2_s (Y)}{q^2_{\bot}} \right)^{1-\lambda_0}
 \,  \ln ( q^2_{\bot} / Q^2_s (Y) )  \,  . 
\label{(4.23)}
\end{equation}
Using (\ref{(4.10)}) we arrive at 
\begin{equation}
\sigma (\vec Q , Y) = x G (  Q , Y) \frac{1}{Q^2} 
\frac{C}{{\tilde C}}
 \,\int  \frac{d^2 q_\bot}{\pi q^2_{\bot}}
\,  H (\vec q_{\bot} , \vec Q)
 \left( \frac{Q^2}{
q^2_{\bot}} \right)^{1-\lambda_0} \,
 \left[ 1 + \frac{\ln (q^2_{\bot} / Q^2)}
{\ln(Q^2 / Q^2_s(Y) )} \right] \,  . 
\label{(4.24)}
\end{equation}
Now we may rewrite, using (\ref{(4.19)}) , QCD factorization as 
given in (\ref{(4.6)}) as 
\begin{equation}
\sigma (\vec Q , Y) = x G ( Q , Y) \frac{1}{Q^2} \int^\infty_0
dy\tilde H (y) \, e^{-2 \bar{\alpha}  \chi (\lambda_0 ) y} \, 
\left[1 + \frac{ \frac{2 \bar{\alpha}  \chi (\lambda_0 ) }{1-\lambda_0} y}{
\ln (Q^2 / Q^2_s (Y))} \right] \,  . 
\label{(4.25)}
\end{equation} 
Using (\ref{(4.14)}) one can write 
(\ref{(4.25)}) as 
\begin{equation}
\sigma (\vec Q , Y) = x G ( Q , Y) \frac{1}{Q^2}\left[  
\tilde H_n - \frac{ \frac{n-1}{1 - \lambda_0} \frac{\partial}{\partial n}}{
\ln (Q^2 / Q^2_s (Y) )} \tilde H_n \right] \Bigg|_{n=n_0} \,  ,
\label{(4.26)}
\end{equation}
with, of course, 
\begin{equation}
\tilde H_n = \int^\infty_0 dy \, {\tilde H} (y) \, e^{-(n-1)y} \, . 
\label{(4.27)}
\end{equation}
Comparing (\ref{(4.24)}) and (\ref{(4.26)}) it is easy to see that they are 
equivalent if 
\begin{equation}
\tilde H_{n_0} = \frac{C}{{\tilde C}} 
\int \frac{d^2 q_\bot}{\pi q^2_{\bot}}
\left( \frac{Q^2}{q^2_{\bot}} \right)^{1-\lambda_0} H (\vec q_{\bot} , \vec Q)
\, ,
\label{(4.28)}
\end{equation}
and
\begin{equation}
- \frac{n_0 - 1}{1 - \lambda_0} \frac{\partial}{\partial n_0} \tilde H_{n_0} =
\frac{C}{{\tilde C}} 
\int \frac{d^2 q_\bot}{\pi q^2_{\bot}} \left( \frac{Q^2}
{q^2_{\bot}} \right)^{1 - \lambda_0} \ln (q^2_{\bot} / Q^2 )
 H (\vec q_{\bot} , \vec Q) \, , 
\label{(4.29)}
\end{equation}
are satisfied. Eq.~(\ref{(4.28)}) is easy to satisfy, if we choose to define 
$\tilde H (y)$ by 
\begin{equation}
\tilde H (y) = \int \, \frac{dn}{2 \pi i} \, e^{(n-1)y} \tilde H_n \, ,
\label{(4.30)}
\end{equation}
and with (\ref{(4.15)})
\begin{equation}
\tilde H_n = \frac{C}{{\tilde C}}
 \int \frac{d^2 q_\bot}{\pi q^2_{\bot} } 
\left( \frac{Q^2}{q^2_{\bot}} \right)^{1-\gamma_n} H (\vec q_{\bot} ,\vec Q ) .
\label{(4.31)}
\end{equation}
Then it is straight forward to see that (\ref{(4.28)}) is satisfied while 
(\ref{(4.29)}) follows by using 
\begin{equation}
\frac{d \gamma_n}{dn} \Bigg|_{n=n_0} = - \frac{(1-\lambda_0)}{n_0 -1} \, ,
\label{(4.32)}
\end{equation}
which, after using (\ref{(4.14)}), is the same as (\ref{(4.22)}). 

Thus we see that $k_\bot$-factorization as expressed in (\ref{(4.5)})
leads to QCD factorization, as expressed in (\ref{(4.6)}), with
$\tilde H (y)$ defined by (\ref{(4.30)}) and (\ref{(4.31)}). While  
$ H (\vec q_{\bot} , \vec Q)$ is a lowest order expression, $\tilde H (y)$ is 
determined in terms of a resummation dictated by (\ref{(4.30)}) and 
(\ref{(4.31)}) and cannot be limited to its lowest order term. The 
simplicity of $k_\bot$-factorization is that all resummations are put into
$\phi_G$ with $H$ remaining a relatively simple quantity, that is the 
hard part defining the reaction is more visible in 
$k_\bot$-factorization than in QCD factorization. 
We remark that including a common additional constant under the logarithms
in (\ref{(4.9)}) and (\ref{(4.10)}), respectively, does not
destroy the derivation given above.
It remains a challenge to 
understand how to extend $k_\bot$-factorization beyond a leading 
logarithmic formalism.


\end{document}